\documentstyle[version2,preprint,aps]{revtex}
\begin{document}
\draft
\begin{title}

Tunneling Conductance Between Parallel Two Dimensional Electron
Systems

\end{title}
\author{Lian Zheng and A.H. MacDonald}
\begin{instit}

Department of Physics,
Indiana University, Bloomington, IN 47405

\end{instit}
\begin{abstract}

We derive and evaluate expressions for the
low temperature {\it dc} equilibrium tunneling
conductance between parallel two-dimensional electron systems.
Our theory is based on a linear-response formalism and
on impurity-averaged perturbation theory.  The
disorder broadening of features in the dependence of
tunneling conductance on sheet densities and in-plane magnetic
field strengths is influenced both by the finite lifetime
of electrons within the wells and by non-momentum-conserving
tunneling events.  Disorder vertex corrections
are important only for weak in-plane magnetic fields and
strong interwell impurity-potential correlations.  We comment
on the basis of our results on the possibility of using
tunneling measurements to determine the lifetime
of electrons in the quantum wells.

\end{abstract}
\narrowtext

\section{Introduction}

 Tunneling of electrons has long been studied
 both as a fundamental manifestation of the quantum nature
 of matter and as a powerful probe of many different
 electronic systems\cite{book}.
 It has only recently become possible to measure the
 tunneling conductance between two-dimensional electron
 systems\cite{ger1,ger2,ger3,jp1,jp2}
 (2DES's) in semiconductors.
 The 2DES's occur in GaAs quantum wells which
 are separated by thin AlAs tunneling barriers \cite{jp1,jp2}.
 The strong constraints imposed by energy and momentum
 conservation in 2D can potentially lead to
 very sharp features in the dependence of the
 equilibrium tunneling conductance
 on the sheet densities in the quantum wells and on the
 strength of an in-plane
 magnetic field.   As we discuss below, tunneling conductance studies
 probe the properties of 2DES's in a way which is
 qualitatively different from conventional in-plane
 transport studies.  In this paper we present a
 theory which provides a framework for the detailed interpretation
 of these experiments and apply it to study the influence of
 disorder on the tunneling conductance.
 We find that these experiments can provide
 a direct measure of the quantum lifetime of electrons
 in the 2DES's.  When combined with measurements of the
 2DES transport lifetime from in-plane transport experiments
 this provides information on the nature of the
 disorder scatterers in the system.   We also show that vertex
 corrections to the naive result for the impurity-broadening
 of features in the tunneling conductance are important
 only for strong correlations between the disorder potentials
 in the separated 2DES's and for weak in-plane magnetic fields.
 In Section II we briefly review the linear response formalism
 which is the basis of our theory.  In Section III we
 present results for the dependence of tunneling
 conductance on the density difference of the 2DES's and
 the strength of an in-plane magnetic field in the absence of
 correlation between the disorder potentials in the two layers.
 In the ideal disorder-free limit the tunnelling conductance
 can be evaluated analytically and
 diverges at certain magnetic field strengths.
 We examine how these divergences are suppressed
 by disorder and discuss the possibility of using the
 resulting tunneling conductance curves to measure the
 quantum lifetime of electrons in the double quantum
 well systems.  In Section IV we examine vertex
 corrections to the naive expression for the tunneling conductance
 in disordered systems.   We find that when the disorder in
 the two 2DES's is strongly correlated features in the
 conductance curves may sharpen.   Vertex corrections are
 important only at weak in-plane magnetic fields.  Finally
 in Section V we briefly summarize our results.

\section{Linear Response Formalism}

We restrict our attention here to the case of zero perpendicular
magnetic field so that states of the 2DES's may be described
in a plane-wave-state basis.   Our interest here is in studying the
dependence of the tunneling conductance on experimentally tunable
parameters like the 2DES layer densities and the strength
of an in-plane magnetic field rather than on its
absolute value.  We therefore describe the tunneling
phenomenologically using a tunneling Hamiltonian
\begin{equation}
\hat H_T = - \sum_{\vec k',\vec k, \sigma} (t_{\vec k', \vec k}
c_{\vec k',\sigma,r}^{\dagger} c_{\vec k,\sigma,l} + h.c.).
\label{equ:1}
\end{equation}
Here we use $ l $ and $ r $ to label states localized in left
and right 2DES's respectively.
The remaining terms in the Hamiltonian may include the interaction
of electrons in either 2DES with a disorder potential and
interactions between electrons in the same or in different
2DES's but are assumed to commute separately
with the total number operator for each 2DES, $ \hat N_{l,r}$.
Following a familiar line\cite{scalapino} we note that the operator
describing the net current flowing from the left
2DES to the right 2DES is given by
\begin{eqnarray}
\hat I &=&-e {d \over dt} \hat N_{r} = e { d \over dt} \hat N_l
 =  i e [ \hat N_r, \hat H_T] / \hbar \nonumber \\
&=& {i e \over \hbar} \sum_{\vec k' \vec k, \sigma}
( t_{\vec k',\vec k} C_{\vec k',\sigma,r}^\dagger C_{\vec k, \sigma, l}
 - h.c.)
\label{equ:2}
\end{eqnarray}
Since the expectation value of $ \hat I $ is zero in any eigenstate
of the Hamiltonian it follows that no current flows in equilibrium.
We evaluate the current which flows in response to a electric
potential difference $V(t) = V \exp (i \omega t) $
between the two wells.  The perturbing term in the Hamiltonian is
\begin{equation}
\hat H_1 (t) = - e V(t) (\hat N_l - \hat N_r) /2.
\label{equ:3}
\end{equation}
Using finite temperature linear response theory we find that
in the limit $ \omega \to 0$
the tunneling conductance, $ G \equiv \langle I \rangle / V$
is given by
\begin{equation}
G = \lim_{\omega \to 0} { \rm{Im} K (\omega + i \eta) \over
\hbar \omega }
\label{equ:4}
\end{equation}
where $ K(z) $ is the analytic continuation of the Matsubara
current-current correlation function:
\begin{equation}
K ( i \omega_n) = - \int_{0}^{\hbar \beta} d \tau
\langle \rm{T}_{\tau} [ \hat I (\tau) \hat I (0) ] \rangle.
\label{equ:5}
\end{equation}
Eq.~(\ref{equ:5}) is the basic expression we use to evaluate the
tunneling conductance.

\section{Uncorrelated Electron Layers}

We first apply Eq.~(\ref{equ:5}) to the case where the
two 2DES's can be considered to be decoupled in the absence
of the perturbation\cite{remark1}.  It then follows from the
translation invariance of each well and the definition of
the Matsubara Greens function \cite{one} that
\begin{eqnarray}
K( i \Omega) &=& { -  2 e^2  \over \hbar^2 \beta }
\sum_{\vec k, \vec k'} \sum_{i \omega_n} | t_{\vec k',\vec k}|^2
 \big[ {\cal G}_{\vec k',r}(i \omega_n) {\cal G}_{\vec k,l} (i \omega_n
 + i \Omega) \nonumber \\
 &+& {\cal G}_{\vec k,r} ( i \omega_n) {\cal G}_{\vec k',l}
 (i \omega_n - i \Omega) \big]
\label{equ:6}
\end{eqnarray}
(We have assumed spin degeneracy.)
The frequency sum may be evaluated by contour integration methods
with the result that
\begin{equation}
G = { 2 e^2 \over \hbar^3}
\sum_{\vec k', \vec k} |t_{\vec k', \vec k}|^2
\int {d \epsilon \over 2 \pi} { - \partial n_F( \epsilon) \over
\partial \epsilon } A_{\vec k', r} (\epsilon / \hbar) A_{\vec k, l}
( \epsilon / \hbar)
\label{equ:7}
\end{equation}
Here $A_{\vec k,m}(\omega) = - 2 \rm{Im}  G_{\vec k,m} ( \omega + i
\eta)$
is the spectral weight of the exact finite temperature Greens function
in layer $m$.  In the absence of interactions and disorder
$A_{\vec k,m} = 2 \pi \hbar \delta( \epsilon - \epsilon_m(\vec k))$
and Eq.~(\ref{equ:7}) reduces to the Fermi golden rule
expression for the tunneling conductance.  Deviations from
this limit probe many-body effects in the 2DES's.
We see from Eq.(~\ref{equ:7}) that when correlation
between the two-layers may be ignored the tunneling experiments
measure the one-body Greens function of the individual layers, i.e.
they measure the spectrum of states created when an extra electron
is added to the ground state of a system.
In contrast in-plane transport experiments measure a two-body
Greens functions of the layers, which reflect the correlations
of particle-hole excitations of a system.  In many-body
systems the information available in these two experiments
is complementary.

The numerical results
we present below are all for the ideal case where the
in-plane component of the momentum is conserved during tunneling,
i.e. where $ t_{\vec k',\vec k} = t \delta_{\vec k',\vec k}$.
Physically this limit corresponds to the case where the tunneling
amplitude is uniform over the area of the 2DES's and the comparison
of our results with experiment\cite{jp1} discussed below
indicates that this uniformity is achieved to a remarkable
degree.  The results presented below
also do not account for electron-electron interactions which
do not appear to play any important role\cite{jp3} experimentally.
We will restrict our attention to $T=0$.

When momentum is conserved during tunneling it follows from
Eq.(~\ref{equ:7}) that in the absence of interactions and
disorder $G \equiv 0$ unless the 2DES's have identical densities.
For this reason it is useful to study the dependence of $G$ on
both the layer densities of the 2DES's and on the strength of
an in-plane magnetic field,
 ${\vec B}=B{\hat e}_y$.   If we neglect
 the finite thickness of the electron layers, the effect of
$\vec B$  on the electronic wavefunctions is just to add a gauge
transformation phase factor which shifts the origin of
momentum space in the two layers.  For example if we
choose the gauge in which the vector potential
is ${\vec A}= Bz{\hat e}_x$, it will shift the origin of
momentum space by $ -z_m {\hat e}_x / \ell^2$
in a layer located in the
plane $z=z_m$.  (Here $\ell = (\hbar c / e B)^{1/2}$ is a magnetic
length.)  With these shifts the single-particle energies in
layer $m$ are given by
\begin{equation}
\epsilon_m(\vec k) = E_m+ {\hbar^2 \over 2m^*} [ (k_x+ z_m/\ell^2)^2 +
k_y^2 ]
\label{equ:8}
\end{equation}
where $E_m$ is the subband energy in layer $m$.  (The layer density
difference is related to the subband energy difference by
$ n_r-n_l =  ( E_l - E_r) / \nu_0$ where $\nu_0 = m / \pi \hbar^2$
is the 2DES density of states.)  In Fig.(~\ref{fig:1}) we show
the tunneling conductance versus in plane-field strength
for the situation
studied experimentally in Ref.(~\cite{jp1}).  The dashed line shows
the result without any disorder which can be obtained
analytically:
\begin{equation}
{G \over A} = {  4 e^2 t^2 m^{*2} \over
    \pi \hbar^5 [ (k_{Fl}+k_{Fr})^2-K^2]^{1/2}[K^2-(k_{Fl}-k_{Fr})^2]^{1/2}}
\label{equ:9}
\end{equation}
when both terms in square brackets are positive and is zero
otherwise.  Here $K = (z_r-z_l)/\ell^2 \sim B$ is the amount by which
the two Fermi circles are displaced as illustrated by the
inset of Fig.(~\ref{fig:1}), $A$ is the area of the
2DES's and $k_{Fm}$ is the Fermi
radius in layer $m$.  Note that for equal densities
in the two layers ($k_{Fl}=k_{Fr}$) $G$ diverges as $B^{-1}$ for
$B \to 0$ but only as $(B_c-B)^{-1/2}$ for $B \to B_c \equiv
\pi |z_r-z_l| /  k_F \Phi_0$.  ($B_c$ is the maximum field
for which tunneling can occur and $\Phi_0$ is the magnetic flux
quantum.)

These divergences of $G$ are suppressed by disorder.
The solid line in Fig.(~\ref{fig:1}) shows results obtained using
the Born approximation expression for the disorder-averaged
Greens function\cite{one,remark2}:
\begin{equation}
G_{\vec k,m}(\omega \pm i \eta) = ( \omega - \hbar^{-1}
\epsilon_m(\vec k) \pm {i \over 2 \tau} )^{-1}.
\label{equ:10}
\end{equation}
Here $\tau$ is the quantum lifetime of the electrons.
The results in Fig.(~\ref{fig:1}) were calculated with
$\tau \epsilon_F / \hbar = 15$.   This level
of disorder is important in suppressing the divergences
at $K=2 k_F $ and especially at $K=0$ but has relatively
little effect on $G$ at other values of $K$.
For equal density 2DES's the maximum conductance reached
at zero in-plane magnetic field ($K=0$) is proportional to
$\tau$:
\begin{equation}
{G(K=0) \over A} = {e^2 t^2 \nu_0 \over
 \hbar^2 \pi } \int d \omega A_B^2(\omega) =
 { 2  e^2 t^2 \nu_0 \tau \over \hbar^2}
\label{equ:11}
\end{equation}
Here $A_B(\omega) = (1/\tau)/(\omega^2+(2/\tau)^2)$ is the
Born approximation Lorentzian spectral function and we have
assumed that $\tau \epsilon_F \gg 1$.
This suggests the possibility of using
the shape of the tunneling conductance curve to measure
the electron lifetime.  For example, assuming the $G(K=k_F)$
is nearly independent of disorder gives
\begin{equation}
{G(K=0) \over G(K=1)} = \sqrt{3}  \tau \epsilon_F / \hbar
\label{equ:12}
\end{equation}
Eq.(~\ref{equ:12}) provides a lower bound on $\tau$ since
the divergence of $G(K=0)$ will also be suppressed by
finite temperatures and also possibly by non-momentum-conserving
tunneling events and macroscopic inhomogeneities.
Another estimate of $\tau \epsilon_F / \hbar$ may be obtained
by examining the dependence of $G$ on the density difference
between the two layers.  In the Born approximation we find that
at $K=0$
\begin{equation}
G(K=0) = {2 e^2 t^2 \nu_0 \tau \over
	  \hbar^2 (1 + 4((n_l-n_r)/(n_l+n_r))^2 (\epsilon_F \tau / \hbar)^2) }
\label{equ:13}
\end{equation}
In Fig.(~\ref{fig:2}) we show results for the layer density
dependence of the conductance at $K=0$ and at several non-zero
values of in-plane magnetic field.  According to
Eq.(~\ref{equ:13}) the relative density difference
at which $G(K=0)$ is reduced to half its equal density
value is $\epsilon_F \tau / \hbar$.

Comparing with the
experimental data of Eisenstein et al.\cite{jp1}
Eq.(~\ref{equ:12}) gives the estimate $\epsilon_F \tau / \hbar
 \sim 5$ while the half-maximum density-difference value
 gives the estimate $\epsilon_F \tau / \hbar \sim 10$ for their
 sample.
These values should be compared with the measured
transport lifetime for the sample of Ref.\cite{jp1},
, $\tau_{tr} \epsilon_F / \hbar \sim 250$.
In high-mobility quantum well structures, scattering is
dominated by ionized
impurities which are spatially separated from the 2DES's.
These impurities produce small-angle-scattering
of electrons because the Coulomb potential they produce is
smooth in the 2DES layers.  Small angle scattering is ineffective
in limiting the transport lifetime.  For
disorder dominated by remote ionized donors large
values of $\tau_{tr}/\tau \sim (k_F d_s)^2 $ are therefore
expected\cite{stern}
and it is widely appreciated that the transport lifetime no longer
effectively characterizes the degree of disorder in the sample.
($d_s$ is the distance between the 2DES and the ionized impurities.)
The estimates of $\tau \epsilon_F / \hbar$ derived from the
tunneling date above are in reasonable accord with expectations,
although the value obtained from Eq.(~\ref{equ:12}) is
probably an underestimate.   The peak value of the conductance
is likely to be reduced by large length scale inhomogeneities in the
samples\cite{jpe3} which we are not adequately described in
the Born approximation.
Any momentum non-conservation during tunneling would also
lead to further broadening of the features in these curves.  The fact
that the $\tau \epsilon_F/ \hbar $ values inferred without accounting
for momentum non-conserving tunneling events are so close to
expected values demonstrates that momentum non-conserving
tunneling events are remarkably infrequent in the
experiments of Ref.(~\cite{jp1}).  This property allows the
tunneling experiments to probe the properties of the 2DES's
in more detail than would otherwise be possible.

\section{Vertex Corrections}

Now we will turn to the discussion of the case where
correlations exist between the disorder potentials
in the two 2DES's.  It is useful to begin with a
formally exact discussion \cite{caveat1}
in terms of the exact eigenstates
in the presence of disorder.
We will see that when the disorder potential in the two layers
are similar,
$V_l({\vec r}) \sim V_r({\vec r})$,
the characteristics of the resonant
tunneling change qualitatively when no magnetic is presented.
The resonant-tunneling
peaks are always sharp, no matter how strong the disorder;
i.e., disorder no longer suppresses the divergence in
the tunneling conductance.

To leading order in |t| we may use the Fermi golden rule
expression for the tunneling conductance\cite{caveat2}($\vec B\ =0$):
\begin{eqnarray}
I&=&{2\pi e\over\hbar}\sum_{\alpha\beta} \vert
t\vert^2\vert\langle\Psi_{\alpha
l}\vert\Psi_{\beta r}\rangle\vert^2\delta (\epsilon_{\alpha l}-
\epsilon_{\beta
r}) \nonumber \\
&\cdot &[\theta (\varepsilon_{f,l}-\varepsilon_{\alpha l})-\theta
(\epsilon_{f,r}-\epsilon_{\beta r})] \\
&= &{2\pi e^2V^{\rm ext}\over\hbar}\sum_{\alpha\beta} \vert
t\vert^2\vert\langle\Psi_{\alpha l}\vert\Psi_{\beta
r}\rangle\vert^2\delta
(\epsilon_{\alpha l}-\epsilon_{\beta r})\delta
(\varepsilon_f-\epsilon_{\alpha l}) \nonumber \label{i}
\end{eqnarray}
where $\Psi_{\alpha l}\ \Psi_{\beta r}$ are the {\it exact}
eigenstates of the
disordered 2DEG systems and the second equality holds to
leading order in the voltage difference between the layers.
When the electrons in different layers have
identical disorder potentials $V_l({\vec r}) \equiv V_r({\vec r})$,
the eigenstates are
orthonormal $\langle\Psi_{\alpha l}\vert\Psi_{\beta r}\rangle
=\delta_{\alpha,\beta}$ \, and $\epsilon_{\alpha l}-\epsilon_{\alpha
r}=E_{ol}-E_{or}$, ($E_{ol}\ E_{or}$ are the energies at the subband edges).
In this case Eq. (\ref{i}) gives the tunneling conductance as:
\begin{eqnarray}
{I\over V^{\rm ext}}&=&{2\pi e^2\over\hbar}\sum_{\alpha} \vert
t\vert^2\delta
(E_{ol}-E_{or})\delta (\epsilon_\alpha -\epsilon_F)\nonumber\\
&=&\left\{\begin{array}{ll}
0 & \mbox{if $E_{ol}\neq E_{or}$}\\
\infty & \mbox{if $E_{ol}=E_{or}$}
\end{array}
\right. \label{ivext}
\end{eqnarray}
A
$\delta$-function like sharp resonant tunneling peak occurs
regardless of the amount of disorder presented, provided that $V_l({\vec
r}) \equiv V_r({\vec r})$.

Clearly, this exact result is not captured in the
discussion of section $V$.  In Feynman diagram language,
we have to sum over diagrams which are important
when the disorder potentials in the two layers are strongly
correlated.  When disorder is treated in the Born approximation
the vertex-correction which yields a conserving
approximation is shown in Fig.(~\ref{fig:3}); it is given by
a sum of ladder diagrams with electron propagators of
opposite 2DES's connected by the rungs of the ladder.
The first diagram in Fig.(~\ref{fig:3}) is the `bubble'
diagram which gives the results of the previous section.

\begin{equation}
S_{\rm RL}(i\omega
)=\sum_{ip_n}
  {\cal G}_{\vec k,r}(ip_n) {\cal G}_{\vec k,l} (ip_n +i \omega_n)
\label{srliw}
\end{equation}
Adding the vertex correction:

\begin{equation}
S_{\rm RL}^{\rm Lad}(i\omega )
={1\over\beta}\sum_{ip_n}{\cal G}_r({\vec p},ip_n){\cal G}_l({\vec
p},ip_n+i\omega )\Gamma_{\vec p}(ip_n,ip_n+i\omega )
\label{srllad}
\end{equation}
where $\Gamma $ satisfies the inhomogeneous integral equation,

\begin{eqnarray}
\Gamma_{\vec K}&(&ip_n,ip_n+i\omega )   \nonumber \\
&=&1+{1\over\hbar^2\nu}\sum_{\vec K'} f_{\rm rl}({\vec k}-{\vec k}')
{\cal G}_r({\vec
K}',ip_n){\cal G}_l({\vec K}',ip_n+i\omega ) \nonumber \\
&\cdot &\Gamma_{{\vec K}'}(ip_n,
ip_n+i\omega )  \label{gammaveck}
\end{eqnarray}
Here the impurity-averaged disorder potential is:
\begin{equation}
\langle V_r({\vec r})V_l({\vec r}')\rangle ={1\over\nu}\sum_{\vec p}
f_{\rm
lr}({\vec p})e^{i{\vec p}({\vec r}-{\vec r}')} \label{vrvecr}
\end{equation}

To illustrate the physics , we consider the following short range
model-potential calculation:
\begin{eqnarray}
f_{rr}({\vec p})&=&f_{ll}({\vec p})=U^2\nonumber\\
f_{lr}({\vec p})&=&\lambda U^2 \label{flr}
\end{eqnarray}
where $U^2 and\ \lambda$ are constants.  The case
of identical impurity potentials in the two 2DES's
corresponds to $\lambda =1$, and the uncorrelated layer limit
in which the `bubble' approximation is appropriate
corresponds to $\lambda =0$.
This simple disorder-potential model
allows us to solve the integral equation [Eq .(\ref{gammaveck})]
very easily, since $\Gamma$ is now independent of wavevector:
\begin{equation}
\Gamma (ip_n,ip_n + i\omega )={1\over 1-{\lambda
U^2\over\hbar^2\nu}\sum_{\vec K} {\cal{G}}_r({\vec K},ip_n)
{\cal{G}}_l({\vec
K},ip_n + i\omega)} \label{gammaminusi}
\end{equation}
With the above expression for $\Gamma$, we can evaluate Eq.
(\ref{srllad})
by contour integration\cite{one}.   After some standard manipulations
we find that the conductance for $\omega\rightarrow 0$ and
$\beta\rightarrow\infty$ (low temperatures) is :
\begin{eqnarray}
G&=&{2t^2 e^2\over \pi{\hbar}^3}
\sum_{\vec p}Re\{ G_r^{\rm adv}({\vec p},0)
G_l^{\rm ret}({\vec p},0)\Gamma_{\vec
p} (-i0^\dagger ,i0^\dagger) \nonumber \\
&-&\ G_r^{\rm adv}({\vec p},0)
G_l^{\rm adv}({\vec p},0)\Gamma_{\vec
p} (i0^\dagger ,i0^\dagger) \}
\label{sumvecp}
\end{eqnarray}

Eq.(\ref{sumvecp}) is easily evaluated numerically and
the results for $ \vec B=0\ $
are shown in Fig.(~\ref{fig:4})   (The second term on the right
hand side becomes negligible compared to the first for
$ \tau \epsilon_F / \hbar \gg 1$.  We see that, for $\lambda
=1$, the resonant tunneling is singularly sharp even for
a finite lifetime; this result is qualitatively different from the
situation  with $\lambda <1$ and demonstrates that the ladder
captures the exact result discussed above.  Numerical results
with finite magnetic fields are shown in Fig.(~\ref{fig:5}); we see that
while vertex corrections still reduce the effects of disorder
the effect is quickly suppressed with increasing field.

In typical double-layer samples the dopant layers are
placed outside of each 2DES.  In this case the disorder in
each layer will come mainly from the closest
dopant layer and we can expect the disorder potentials
in the two layers to be weakly correlated.  However in
some circumstances it may be desirable to place the doping
layer in the middle of the barrier separating 2DES's.
In this case the disorder potentials in the two layers
will be strongly correlated.  The calculations in this
section show that, even though doping in the center
of the barrier is likely to decrease $\tau$ substantially
it will not effect tunneling between the 2DES's at $B=0$.~\cite{rk3}

\section{ summary}

In summary we have derived and evaluated expressions for the
low temperature {\it dc} equilibrium tunneling
conductance between parallel two-dimensional electron systems.
Analytic results were obtained for the dependence of the
tunneling conductance on the strength of an in-plane
magnetic field and on the layer density difference between
the 2DES's.  The effect of uncorrelated disorder on the
tunneling conductance was discussed.  Expression relating
measures of the broadening of conductance curves
by disorder to $\tau \epsilon_F / \hbar$ were given and
the possibility of using them to measure the electron
lifetime in a 2DES was discussed.  Comparisons of our
theoretical results to the experimental results of
Eisenstein {\it et al.} demonstrate that non-momentum-conserving
tunneling events are remarkably infrequent in their samples.
Finally we have shown that vertex corrections become important,
especially at zero in-plane field, when strong correlations
exist between the disorder potentials in the two 2DES's.
These vertex corrections are likely to be important in
practice in samples where the doping layer is in the middle of
the barrier separating the 2DES's.

\acknowledgments

The authors are grateful for many stimulating and informative
interactions with J.P. Eisenstein.  Helpful conversations
with G.S. Boebinger, Jun Hu and Eric Yang are also acknowledged.
This work was supported by the National Science Foundation under
grant DMR-9113911.

\newpage

\figure{ Tunneling conductance {\it vs.} in-plane
magnetic field strength at equal layer density,
$n_l=n_r=1.5 \times 10^11 {\rm cm}^{-2}$.
The insert shows the displaced Fermi circles in the
two layers.  Momentum conserving tunneling events at the
Fermi energy are allowed at the intersections of the two
Fermi circles.  The dashed line shows the result
in the absence of disorder while the solid line was
evaluated for a quantum lifetime $\tau = 15 \hbar / \epsilon_F$.
$G_0/A = 1.5\times 1.0^2(m^{*2} e^2 t^2 / \pi^4 \hbar^5 n_l)$.
\label{fig:1}}

\figure{ Tunneling conductance {\it vs.} the ratio of electron
densities in the two layers.   ($n_l = 0.9 \times 10^{11}$cm$^{-2}$
These results were calculated with
$\tau \epsilon_F / \hbar =15 $.   The dashed line shows the
result obtained in the absence of disorder at $B=0.4$Tesla.
$G_0/A = 0.9\times 1.0^2(m^{*2} e^2 t^2 / \pi^4 \hbar^5 n_l)$.
\label{fig:2}}

\figure{Ladder diagrams for the vertex correction approximation
for the current current response function.
\label{fig:3}}

\figure{Tunneling conductance versus layer density ratios
at zero in-plane magnetic field
for different degrees of correlation between the impurity
potentials in the two 2DES's.  As the correlation becomes
strong ($\lambda \to 1$) tunnelling occurs only at nearly
equal densities even if there is substantial disorder in each layer.
\label{fig:4}}

\figure{Tunneling conductance versus layer density ratios
at in-plane magnetic field $\vec B =0.4T$
for different degrees of correlation between the impurity
potentials in the two 2DES's.   The correlation does
not change the tunnelling behavior qualitatively.
\label{fig:5}}


\begin{thebibliography}{13}
\bibitem{remark1}  This will be the case if tunnelling,
interlayer electron-electron interactions, and correlation
between disorder potentials in opposite wells can be neglected.
\bibitem{remark2} We make a common approximation by ignoring
the contribution of disorder to the real part of the self-energy.
This is justified when the disorder is sufficiently weak as is
case in high-mobility samples.  In the Born approximation
$\tau$ is the Fermi-golden-rule lifetime for an electron prepared
in a state of definite wavevector.
\bibitem{caveat1} This discussion is exact for the idealized
system of non-interacting electrons.
\bibitem{caveat2} This discussion can be generalized to
the case of larger $|t|$.
\bibitem{rk3} Doping in the center of the barrier may increase the
importance of non-momentum-conserving tunneling, however,
and obscure the effect discussed in this section.
\end{thebibliography}
\end{document}